# Quantum control of hybrid nuclear-electronic qubits

Gavin W. Morley[1,2,†], Petra Lueders[3], M. Hamed Mohammady[1], Setrak J. Balian[1], Gabriel Aeppli[1,2], Christopher W. M. Kay[2,4], Wayne M. Witzel[5], Gunnar Jeschke[3] & Tania S. Monteiro[1]

1 Department of Physics and Astronomy, University College London, Gower Street, London WC1E 6BT, UK

2 London Centre for Nanotechnology, University College London, Gordon Street, London WC1H 0AH, UK

3 Laboratory of Physical Chemistry, ETH Zurich, Wolfgang-Pauli-Str. 10, 8093 Zurich, Switzerland

4 Institute of Structural and Molecular Biology, University College London, Gower Street, London WC1E 6BT, UK

5 Sandia National Laboratories, Albuquerque, New Mexico 87185, USA

[†] Present address: Department of Physics, University of Warwick, Gibbet Hill Road, Coventry CV4 7AL, UK

Pulsed magnetic resonance is a wide-reaching technology allowing the quantum state of electronic and nuclear spins to be controlled on the timescale of nanoseconds and microseconds respectively[1]. The time required to flip either dilute electronic or nuclear spins is orders of magnitude shorter than their decoherence times[2-8], leading to several schemes for quantum information processing with spin qubits[9-12]. We investigate instead the novel regime where the eigenstates approximate 50:50 superpositions of the electronic and nuclear spin states forming "hybrid nuclear-electronic" qubits[13,14]. Here we demonstrate quantum control of these states for the first time, using bismuth-doped silicon, in just 32 ns: this is orders of magnitude faster than previous experiments where pure nuclear states were used[4,5]. The coherence times of our states are five orders of magnitude longer, reaching 4 ms, and are limited by the naturally-occurring $^{29}$Si nuclear spin impurities. There is quantitative agreement between our experiments and no-free-parameter analytical theory for the resonance positions, as well as their relative intensities and relative Rabi oscillation frequencies. In experiments where the slow manipulation of some of the qubits is the rate limiting step, quantum computations would benefit from faster operation in the hybrid regime.

Phosphorus atoms are used to increase the conductivity of silicon because they donate their excess valence electron to the silicon conduction band. At low temperatures, and sufficiently low phosphorus concentrations, these electrons become bound to phosphorus nuclei and their spins can be used to store quantum information[2,3,6,7,15,16]. Electronic readouts have increased the sensitivity of these measurements[15,17] and Rydberg states can be coherently controlled also[18].

Bismuth is in the same column of the periodic table as phosphorus, but because it plays no part in the semiconductor industry it has been overlooked as a quantum bit host until recently[4,5,13,14,19,20]. Early electron paramagnetic resonance (EPR) spectroscopy[21,22] found that bismuth in silicon (Si:Bi) has a very large electron-nuclear hyperfine coupling of $A/2\pi = 1.4754$ GHz. Combining this with the large nuclear spin of $I = 9/2$ allows us to work in the hybrid nuclear-electronic regime where the electron-nuclear coupling is comparable to the electronic Zeeman energy for magnetic fields within the range of pulsed EPR spectrometers. The coupled spin Hamiltonian for Si:Bi is otherwise identical to that for the well-studied Si:P. It takes the form:

$$\hat{H}_0 = \omega_0 \hat{S}_z - \omega_0 \delta \hat{I}_z + A\hat{\mathbf{S}}.\hat{\mathbf{I}} \qquad (1)$$

where the electronic Zeeman frequency, $\omega_0 = g\mu_B B/\hbar$, depends on the Dirac constant ($\hbar$), the applied magnetic field ($B$), the Bohr magneton ($\mu_B$), and the Landé $g$-factor which is $g = 2.0003$ for Si:Bi. The nuclear Zeeman frequency is $\omega_0 \delta$ with $\delta = 2.488 \times 10^{-4}$. Strong hyperfine coupling has previously been shown to induce interesting effects, such as forestalling an electronic quantum phase transition in a heavy rare-earth ion[23]. Here we show that strong hyperfine coupling can provide a useful new regime for spin qubits.

Figure 1a shows the energy-level spectrum of Si:Bi as a function of magnetic field, $B$. There are 20 energy levels. In the high-$B$ limit these correspond to good quantum numbers for the electronic $m_s = \pm 1/2$ and nuclear $-9/2 \leq m_I \leq 9/2$ spin states. The zero-field limit also yields two good quantum numbers $F = S + I$ and $m = m_s + m_I$. Since $[\hat{H}_0, \hat{F}_z] = 0$, the total $m$ remains a good quantum number throughout, even in the intermediate field regime ($B \approx 0.05 - 0.5$ T), which is of special interest.

The most notable feature in Fig. 1a is a chain of four equally-spaced (and overlapping) avoided crossings. The four pairs of quantum states involved each become a maximal mix of the electronic and nuclear degrees of freedom at the crossing: in the $|m_s, m_I\rangle$ basis they adopt a "Bell-like" form: $|\Psi^\pm\rangle = \frac{1}{\sqrt{2}}\left[|\frac{1}{2}, m-\frac{1}{2}\rangle \pm |\frac{-1}{2}, m+\frac{1}{2}\rangle\right]$ where the concurrence reaches a maximal value of 1. As the crossings are broad and overlapping, all eight states will be substantially mixed as shown with the concurrence colour-scale in Fig. 1a. We term these our hybrid nuclear-electronic qubit states. States $|10\rangle$ and $|20\rangle$ in Fig. 1a remain unmixed throughout; previous work[19] has demonstrated hyperpolarization of over 90% to state $|10\rangle$, suggesting a computational realization may begin by initializing onto this state.

Previously-performed Si:Bi magnetic resonance, with excitation above $5A = 7.4$ GHz, found ten resonances which become pure EPR transitions at high magnetic fields[4], obeying the EPR selection rules $\Delta m_s = \pm 1, \Delta m_I = 0$. We access spin dynamics in the hybrid nuclear-electronic regime below 7.4 GHz (see Supplementary Fig. S3) using a custom-built pulsed 4 GHz spectrometer at ETH Zurich[24].

We first explain briefly the intermediate-field behaviour. In ref. 13 we showed that the Si:Bi quantum spin states involved in mixing correspond to doublet states with constant $m$. The Hamiltonian factors into block-diagonal form with 2×2 sub-Hamiltonians, $h_m$, each corresponding to one of the 9 doublets, $-4 < m < +4$ where:

$$h_m = \Delta_m \hat{\sigma}_z + \Omega_m \hat{\sigma}_x - \varepsilon_m \mathbb{1} \tag{2}$$

with $\Delta_m = \tfrac{1}{2}[mA + \omega_0(1+\delta)]$, $\Omega_m = \tfrac{A}{2}(25-m^2)^{1/2}$, $\varepsilon_m = \tfrac{A}{4} + m\delta\omega_0$ and the identity operator $\mathbb{1}$. The Pauli matrices, $\hat{\sigma}_z$ and $\hat{\sigma}_x$, do not correspond to a physical spin, but simply to the 2×2 dimension of the Hamiltonian coupling. Nevertheless we can consider the doublets to be eigenstates of an effective pseudo-spin $\hat{\sigma}_n = \hat{n}_m \cdot \hat{\sigma}$ by re-writing the Hamiltonian in equivalent form:

$$h_m = \beta_m \hat{n}_m \cdot \hat{\sigma} - \varepsilon_m \mathbb{1} \tag{3}$$

where the unit vector, $\hat{n}_m = (\sin\theta_m, 0, \cos\theta_m)$. Hence $\beta_m^2 = \Delta_m^2 + \Omega_m^2$, while $\theta_m = \arctan\Omega_m/\Delta_m$. Thus the eigenenergies of the doublets are given analytically in terms of the external magnetic field $\omega_0$, the hyperfine coupling $A$, and the nuclear Zeeman correction $\delta$. The eigenenergies of the doublets are simply $E_m^\pm = \pm\beta_m - \varepsilon_m$, while the eigenstates are

$$|\pm, m\rangle = \cos\frac{\theta_m}{2}\left|\pm\frac{1}{2}, m\mp\frac{1}{2}\right\rangle \pm \sin\frac{\theta_m}{2}\left|\mp\frac{1}{2}, m\pm\frac{1}{2}\right\rangle. \tag{4}$$

The eigenstates will take a Bell-like form whenever $\cos(\theta_m/2) = \sin(\theta_m/2) = 1/\sqrt{2}$; this occurs when $\theta_m = \pi/2$, which corresponds to $\Delta_m = \tfrac{1}{2}[mA + \omega_0(1+\delta)] = 0$ and thus $-mA \approx \omega_0$. The condition represents a series of points where the hyperfine interaction cancels the electronic Zeeman splitting. For the doublets, this occurs for $-\omega_0/A \approx m = -1, -2, -3, -4$ (excluding the trivial $\omega_0 = 0$, where selective excitation of individual transitions is impossible) and accounts for the four equally-spaced avoided crossings. At larger magnetic fields, $\theta_m \to 0$, the second term in Eq. 4 vanishes and the states become unhybridized.

States $|10\rangle$ and $|20\rangle$ are not part of a constant-$m$ doublet; they always remain unmixed and correspond to $m = -5$ and $m = +5$ respectively. In effect, $\theta_{\pm 5} = 0$ for all fields and the corresponding eigenenergies for these states are $E_{m=\pm 5} = \pm\frac{1}{2}\omega_0 \mp \frac{9}{2}\omega_0\delta + \frac{9A}{4}$. For all magnetic fields we can write $|10\rangle = |m_s = -1/2, m_I = -9/2\rangle$.

Magnetic resonance couples states in adjacent doublets, thus corresponding to $m \to m \pm 1$. From the above expressions, we find that, with 4.044 GHz excitation, two transitions appear, at $B$ = 145.6 mT (transition 10-9) and at $B$ = 345.0 mT (transition 11-10); these are shown in Fig. 1b as stars. In the high-field (unhybridized) limit the corresponding spin flips would be:

$$11 \to 10 \equiv |m_s = \tfrac{1}{2}, m_I = -\tfrac{9}{2}\rangle \to |-\tfrac{1}{2}, -\tfrac{9}{2}\rangle$$
$$10 \to 9 \equiv |m_s = -\tfrac{1}{2}, m_I = -\tfrac{9}{2}\rangle \to |-\tfrac{1}{2}, -\tfrac{7}{2}\rangle. \tag{5}$$

The $10 \to 9$ transition violates the selection rule for electron spin transitions so would not be seen if the nuclear and electronic states were not mixed. The states $|11\rangle$ and $|9\rangle$ comprise the $m = -4$ doublet, so from the preceding equations, we see that for $B$ = 0.15 T, $\theta_{-4}$ = 0.62π, while for $B$ = 0.35 T, $\theta_{-4}$ = 0.28π. For the former, the transition strength $\langle 10|S_x|9\rangle \propto \sin(\theta_{-4}/2)$ while for the latter, the transition strength $\langle 11|S_x|10\rangle \propto \cos(\theta_{-4}/2)$. Substituting for the angles, we see that the ratio of the transition strengths $\langle 11|S_x|10\rangle / \langle 10|S_x|9\rangle = 1.1$. The Rabi oscillation rates scale with transition strengths, so this prediction is strikingly different to magnetic resonance experiments performed to date (in the high-field limit) where Rabi oscillations take around $10^3$ times longer for NMR than for EPR[1]. The expected ratio of the magnetic resonance intensities is then $1.1^2 \approx$ 1.2.

Figure 1b depicts the continuous-wave magnetic resonance spectrum of Si:Bi at 4 GHz, measured at a temperature of 42 K. Magnetic field modulation was used resulting in the differential lineshapes. The positions and relative intensities of the two Si:Bi hybrid nuclear-electronic resonances agree with analytical predictions with no free parameters using EPR measurements at 9.7 GHz (refs. 4, 13). 42 K was used because at lower temperatures than this, the long spin relaxation times of Si:Bi lead to saturated (bleached) spectra even with relatively low microwave power. At 42 K, some of the bismuth atoms donate their electrons to the conduction band, and these mobile electrons reduce the Si:Bi spin-lattice relaxation time enough that the resonances are not saturated by the excitation. The conduction electrons absorb microwaves not through magnetic resonance, but via the microwave *electric* field, with a very broad "Drude" resonance centred on zero magnetic field. The Drude resonance was subtracted here, but can be seen in the Supplementary Information as a negative background slope in Fig. S1a. At temperatures above ~42 K, the Drude absorption is so large that the microwave resonator with the sample within does not resonate, preventing measurements.

The intensity of the 11-10 resonance is slightly larger than the 10-9 resonance as expected. The resonances are fitted with differential Gaussian peaks in Fig. 1b. Integrating these produces the Gaussian peaks shown in Supplementary Fig. S1b, and the area under these Gaussians is proportional, among other things, to the number of spins present. The measured ratio is $area_{11-10}/area_{10-9} = 1.2$, in agreement with the value predicted above.

It has previously been shown that CW spectra can be recorded for Si:P in the regime where the hyperfine energy is comparable to the Zeeman energy, but this required very low magnetic fields below 20 mT and no quantum control or spin dynamics were reported[26]. Here we go on to demonstrate coherent quantum control of the Si:Bi spin states in the hybrid nuclear-electronic regime.

Figure 2a shows Rabi oscillations for both of the 4 GHz Si:Bi resonances at 8 K with the same microwave power. Fourier transforming the data reveals that the time for a $\pi$ pulse is around 32 ns for the microwave power (~1 kW) used here. The ratio of the 11-10 Rabi frequency to the 10-9 is measured as 1.1, in agreement with the ratio of the predicted transition strengths described above. The 10-9 resonance would correspond to an NMR experiment in the high-field regime, requiring orders of magnitude longer pulses. Fast control of nuclear spin states has been demonstrated previously[27] with microwave pulses by aligning a crystal of malonic acid so as to reach the "exact cancellation" condition[1]. Unfortunately the coherence times of electron spins in malonic acid only reach 30 μs[28].

Fig. 3 shows that the coherence times of Si:Bi hybrid nuclear-electronic spins can reach 4 ms. These were obtained simply by recording the decay of Hahn echoes as the two excitation pulses were separated; example decay curves are shown in Fig. 3a and in the Supplementary Information. For temperatures above ~14 K, the decay curve is mono-exponential and limited by the $T_1$ spin-lattice relaxation. For lower temperatures, the spin echo decay does not track the $T_1$, and becomes non-exponential. We fitted these decays with the function $\exp(-t/T_2 - t^n/T_S^n)$ as used previously[2-6], where the time from the first pulse to the echo is $t$, and the exponent $n$ is commonly between 2 and 3 for Si:P and Si:Bi. $T_S$ characterizes the decoherence time due to naturally-occurring $^{29}$Si impurities[29,30], which could be removed in future samples[2,3,6].

Electronic[6] and nuclear[7] decoherence times have both been measured in $^{28}$Si:P to be up to 10 seconds. If the same behaviour were found for $^{28}$Si:Bi then the hybrid nuclear-electronic regime studied here would offer faster control of the qubits with coherence times that are at least as long as in the unhybridized regime. However, it is generally expected that for $^{28}$Si:P and $^{28}$Si:Bi in the unhybridized regime, it should be possible to measure longer spin coherence times for the nuclear spins than the electron spins because the smaller nuclear magnetic moment is more weakly coupled to its environment. For a computation making use of both the electronic and nuclear spins as qubits, the time available for quantum computing is the shortest coherence time: that of the electron spins. Conversely, the total time taken for the quantum computation is dominated by the time needed for the slowest manipulation: that of the nuclear spins. The hybrid regime would then provide a big advantage for quantum computing because the limiting coherence time is at least as long, while quantum control is orders of magnitude faster.

We have demonstrated here a speed-up of more than two orders of magnitude, from 4 μs to 32 ns, but these times depend on the power of the driving electromagnetic radiation used. The underlying speed-up is the ratio of the electronic magnetic moment to the nuclear magnetic moment, $1/\delta = 4000$. If 'quantum memory'[7] were desired rather than quantum information processing, the unhybridized regime may be better, in which case it could be useful to be able to switch between the two regimes with a magnetic field jump, as discussed in the Supplementary Information.

The decoherence times for Si:Bi we measure at 4 GHz are similar to those measured previously[4] for the same sample at 10 GHz, as shown in Fig. 3b. We attribute this to the similar gradient $df/dB$ in the two cases, where $f$ is the resonant frequency[5,13,14] (see Supplementary Fig. S3).

In conclusion, we have experimentally demonstrated quantum dynamics of a hybrid nuclear-electronic spin system with coherence times five orders of magnitude longer than the timescale for manipulation. This manipulation consists of 32 ns spin flips that would require microseconds if the external magnetic field were not comparable to the hyperfine interaction.

**Acknowledgements** We acknowledge Bernard Pajot for the Si:Bi crystal used here, René Tschaggelar for technical assistance, the National EPR Facility & Service at the University of Manchester, UK, for initial continuous-wave experiments at 4 GHz and the EPSRC COMPASSS grant. Sandia National Laboratories is a multiprogram laboratory operated by Sandia Corporation, a wholly owned subsidiary of Lockheed Martin Corporation, for the U.S. Department of Energy's National Nuclear Security Administration under contract DE-AC04-94AL85000. G.W.M. is supported by an 1851 Research Fellowship.

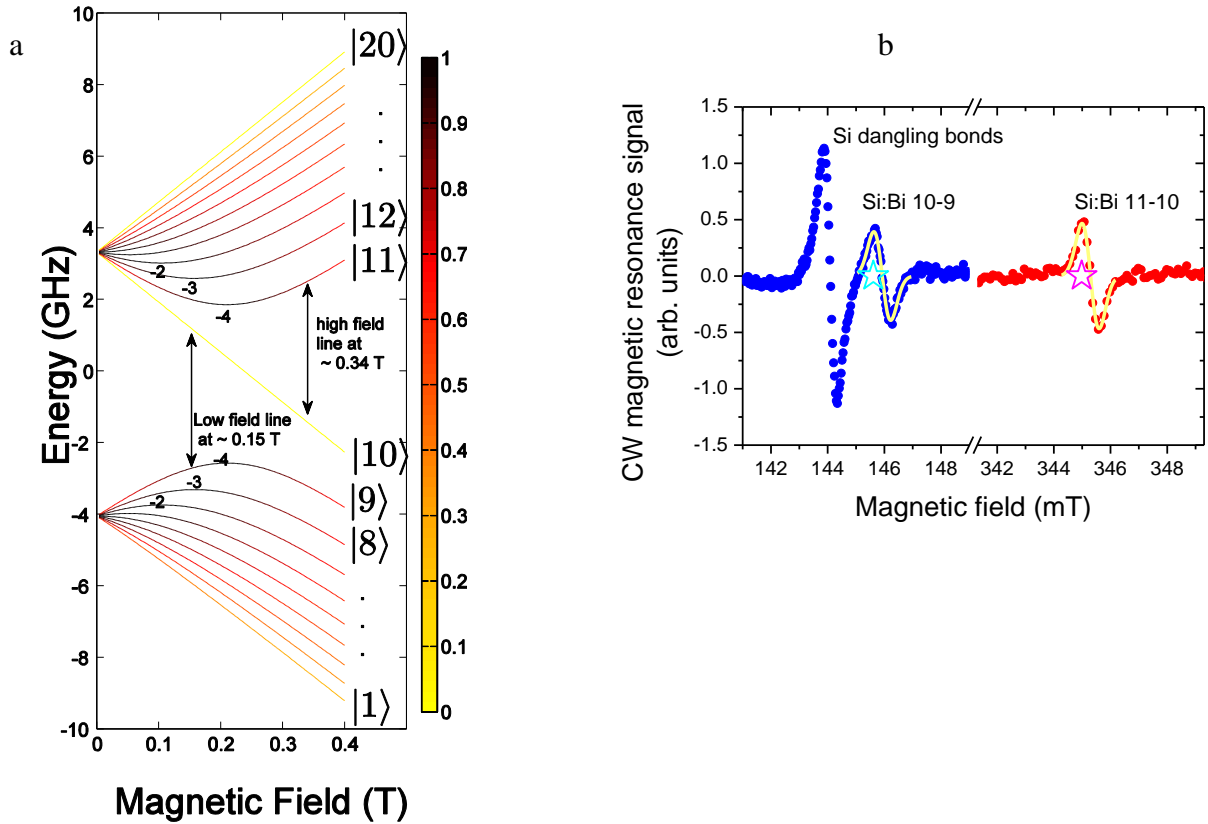

**Figure 1. Spectroscopy of hybrid nuclear-electronic qubits.** a) Analytical simulation of the 20 Si:Bi energy levels. The electron-nuclear entanglement is shown as a colour-scale using concurrence = 0 (no entanglement, yellow) to 1 (maximum entanglement, black). The two transitions that exist for 4 GHz excitation are shown with vertical arrows, and they link state 10 (which is always unentangled) to states with large concurrences of 0.92 (state 9) and 0.76 (state 11). The Supplementary Information shows that magnetic field jumps could be used to move from a regime where the electronic and nuclear spins are not good quantum numbers to a regime where they are. b) Magnetic resonance spectrum of bismuth-doped silicon (Si:Bi) at 42 K with 4.044 GHz continuous-wave (CW) excitation. The solid lines are fits with differentiated Gaussians having full-width-half-maximum of 0.7 mT. The predicted positions of both Si:Bi resonances are shown as stars. The silicon dangling bonds around 144 mT are $P_b$ centres[25] at the interface of Si and $SiO_2$. The signal from the conduction electrons has been subtracted as shown in the Supplementary Information. The sample is a single float-zone crystal of silicon, bulk-doped in the melt with bismuth atoms to ~$3 \times 10^{15}$ Bi $cm^{-3}$. We used the same sample previously for measurements at 10 GHz and 240 GHz[4,13].

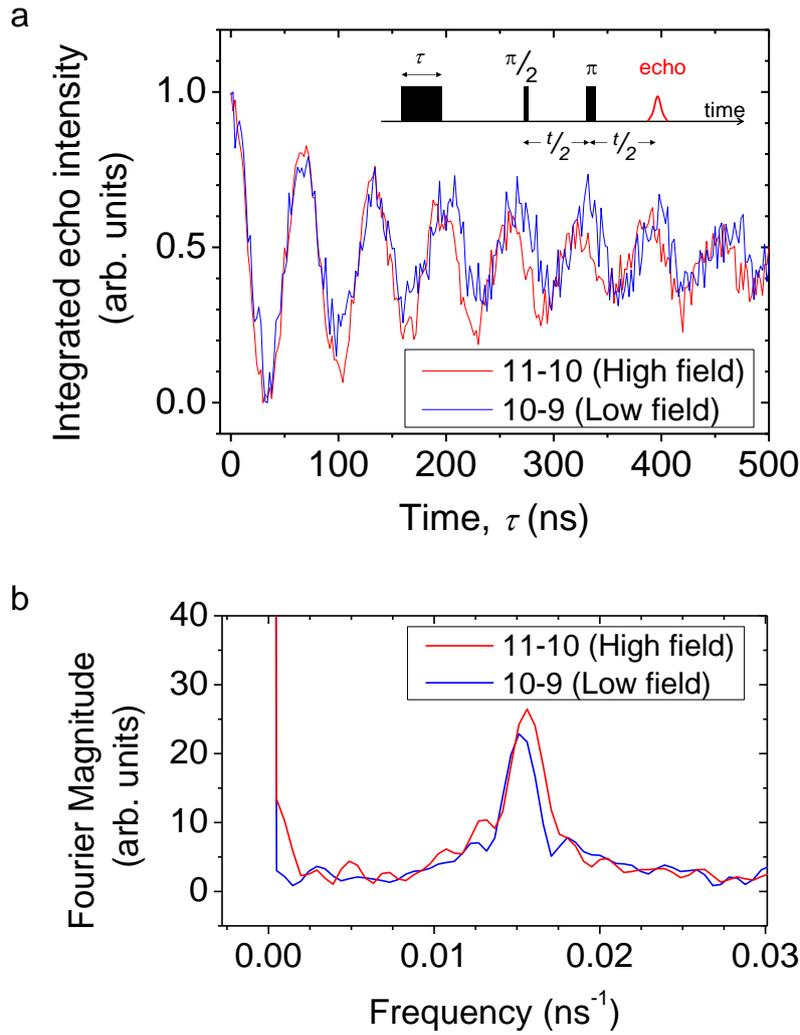

**Figure 2. Fast quantum control of hybrid nuclear-electronic qubits.** a) Rabi oscillations demonstrate coherent control of both of the 4 GHz hybrid nuclear-electronic transitions. At higher magnetic fields, the 11-10 resonance becomes an EPR transition, while the 10-9 resonance becomes an NMR transition. Controlling this NMR transition in the past[4,5] has required π pulses of ≥4 μs, two orders of magnitude longer than the 32 ns π pulses we use here. b) Fourier transforming the Rabi oscillations reveals that the 11-10 transition experiences 10% faster nutation as expected.

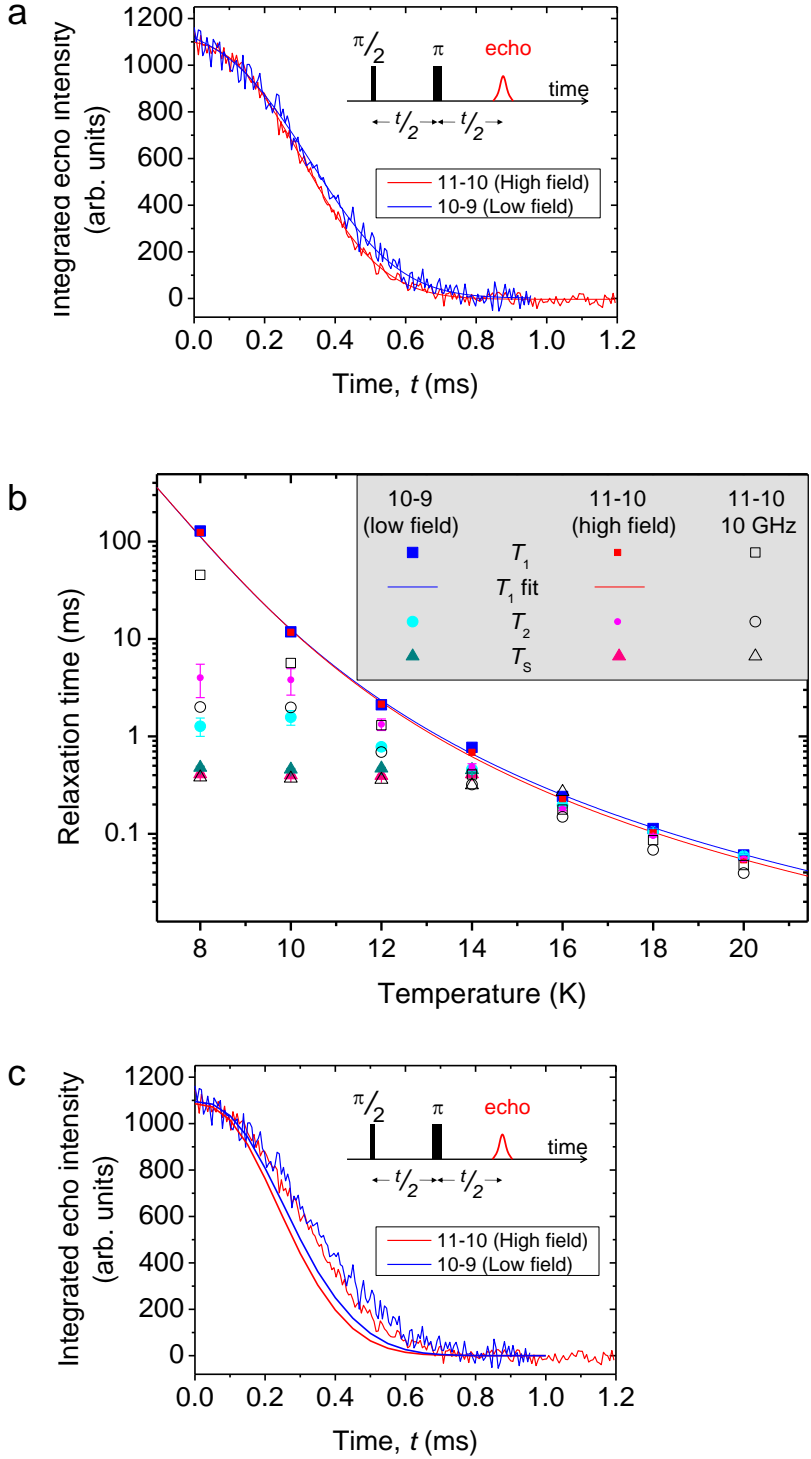

**Figure 3. Coherence times of hybrid nuclear-electronic qubits.** a) Example spin echo coherence decays measured for both Si:Bi transitions at 4 GHz, with a temperature of 10 K. The echo coherence decay is limited by $^{29}$Si nuclear spins, as parameterized by $T_S$ in the fitting function $\exp(-t/T_2 - t^n/T_S^n)$, where $T_2$ is the spin coherence time. The exponent $n$ was used as a fitting parameter as discussed in the Supplementary Information. b) Nuclear-electronic spin relaxation times as a function of temperature for both resonances at 4 GHz with previously published data[4] at 10 GHz for comparison. The 4 GHz spin-lattice relaxation rates, $T_1^{-1}$, were fitted[22] with a power law (Raman) term added to an exponential (Orbach) term using the formula $T_1^{-1} = PT^7 + E\exp(-\Delta/k_BT)$ for temperature $T$ where $k_B$ is the Boltzmann constant. The fitting parameters take the values (with standard errors) $P_{11\text{-}10} = 2.7 \pm 0.7 \times 10^{-6}$ s$^{-1}$K$^{-7}$, $P_{10\text{-}9} = 2.6 \pm 1 \times 10^{-6}$ s$^{-1}$K$^{-7}$, $E_{11\text{-}10} = 4 \pm 0.5 \times 10^6$ s$^{-1}$, $E_{10\text{-}9} = 3 \pm 0.7 \times 10^6$ s$^{-1}$, $\Delta_{11\text{-}10}/k_B = 113 \pm 3$ K and $\Delta_{10\text{-}9}/k_B = 110 \pm 4$ K for the two Si:Bi transitions. The magnetic field was perpendicular to the [111] direction of the crystal for all of the measurements presented in this paper. The temperature was controlled to better than ±0.05 K. All of the pulsed measurements presented in this paper used 16 ns $\pi/2$ pulses and 32 ns $\pi$ pulses with two-step phase cycling[1]. Phase-noise is not visible in the Hahn echo decays despite the fact that we did not implement "magnitude detection"[1-6]. The $T_1$ data in Fig. 3b were recorded with the standard inversion recovery pulse sequence[1]. The error bars in Fig. 3b show the standard errors, which are in many cases smaller than the symbol. c) The same data as in Fig 3a, but instead of fits, the smooth curves show a simulation with no free parameters using the cluster correlation expansion[30]. This demonstrates that the $^{29}$Si impurities are the dominant contribution to the decay of the spin coherence. A 27.8 nm cube was used for the lattice and clusters with two $^{29}$Si spaced by up to the third nearest neighbour distance were included. The thickness of the line is of the order of the standard deviation of the mean intensities after 100 random spatial configurations of $^{29}$Si nuclei. The Supplementary Information shows that the expansion converges as the lattice size and the distance between paired $^{29}$Si nuclei are increased.

# Supplementary Information

# Quantum control of hybrid nuclear-electronic qubits

Gavin W. Morley[1,2,†], Petra Lueders[3], M. Hamed Mohammady[1], Setrak J. Balian[1], Gabriel Aeppli[1,2], Christopher W. M. Kay[2,4], Wayne M. Witzel[5], Gunnar Jeschke[3] & Tania S. Monteiro[1]

[1] Department of Physics and Astronomy, University College London, Gower Street, London WC1E 6BT, UK
[2] London Centre for Nanotechnology, University College London, Gordon Street, London WC1H 0AH, UK
[3] Laboratory of Physical Chemistry, ETH Zurich, Wolfgang-Pauli-Str. 10, 8093 Zurich, Switzerland
[4] Institute of Structural and Molecular Biology, University College London, Gower Street, London WC1E 6BT, UK
[5] Sandia National Laboratories, Albuquerque, New Mexico 87185, USA
[†] Current address: Department of Physics, University of Warwick, Gibbet Hill Road, Coventry CV4 7AL, UK, gavin.morley@warwick.ac.uk

## Si:Bi spectra

Figure S1a shows the continuous-wave (CW) spectrum before subtraction of the background signal due to conduction electrons. The fitted Gaussian linewidth (full width at half maximum with standard errors) is 0.70 ±0.005 mT for the 10-9 resonance and 0.71 ±0.01 mT for the 11-10 resonance. We also measured linewidths from 0.55 − 0.70 mT with pulsed 4 GHz measurements at lower temperatures as shown in Fig. S2. These values are larger than the 0.4 − 0.5 mT linewidths[4] measured at 10 GHz. The EPR linewidth of dilute donors in silicon is due to inhomogeneous broadening by $^{29}$Si impurities[4,5,21,29,30]. The large value of 0.7 mT at 42 K may result from broadening due to coupling with the very broad conduction electron signal.

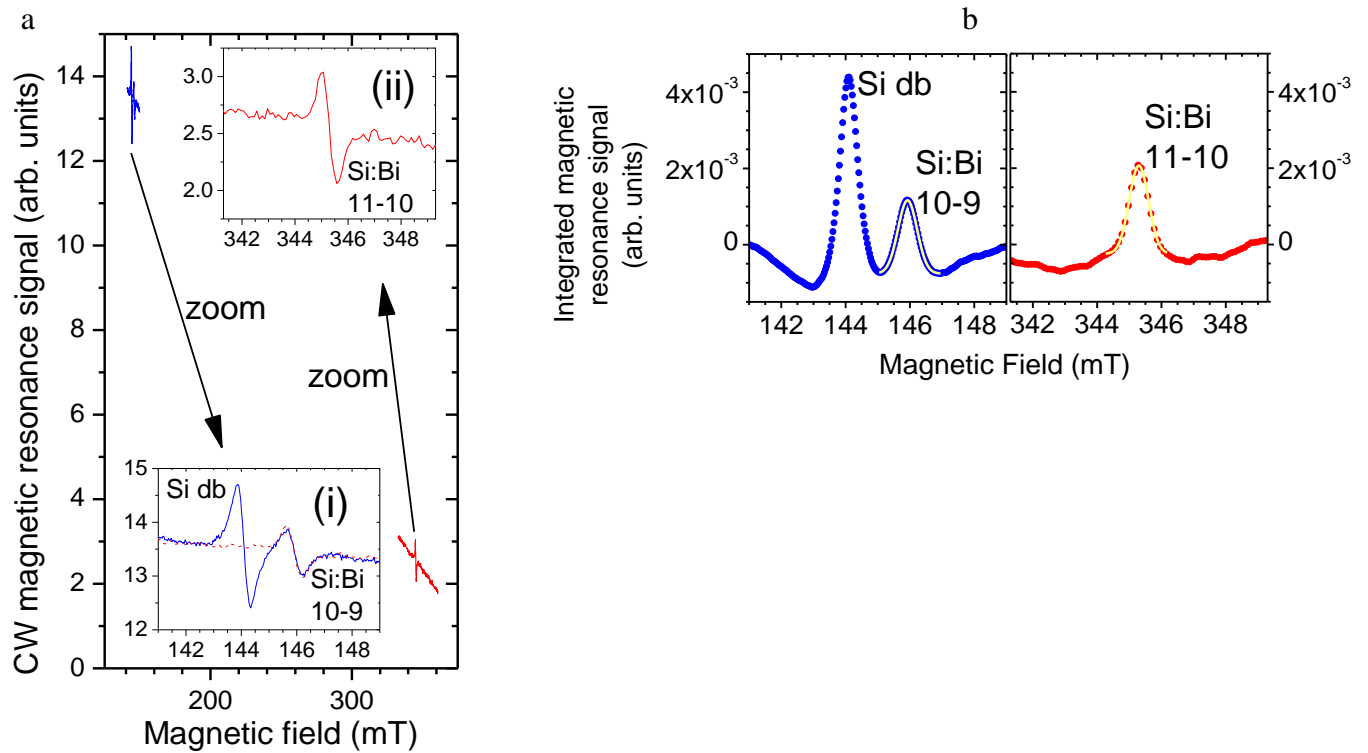

**Figure S1. Magnetic resonance spectrum of Si:Bi at 4 GHz.** a) The same data as in Fig. 1b of the main paper, before subtraction of linear fits to remove the broad background signal from conduction electrons. Shifting the 11-10 Si:Bi resonance up and to the left produces the dotted red line in inset (i), showing that it is very similar in size and shape to the 10-9 transition. b) Integrating the spectra after subtraction of the broad conduction electron signal produces Gaussian peaks. The areas under these Gaussians are in the ratio area$_{11\text{-}10}$/area$_{10\text{-}9}$ = 1.2, in agreement with the theory in the main text.

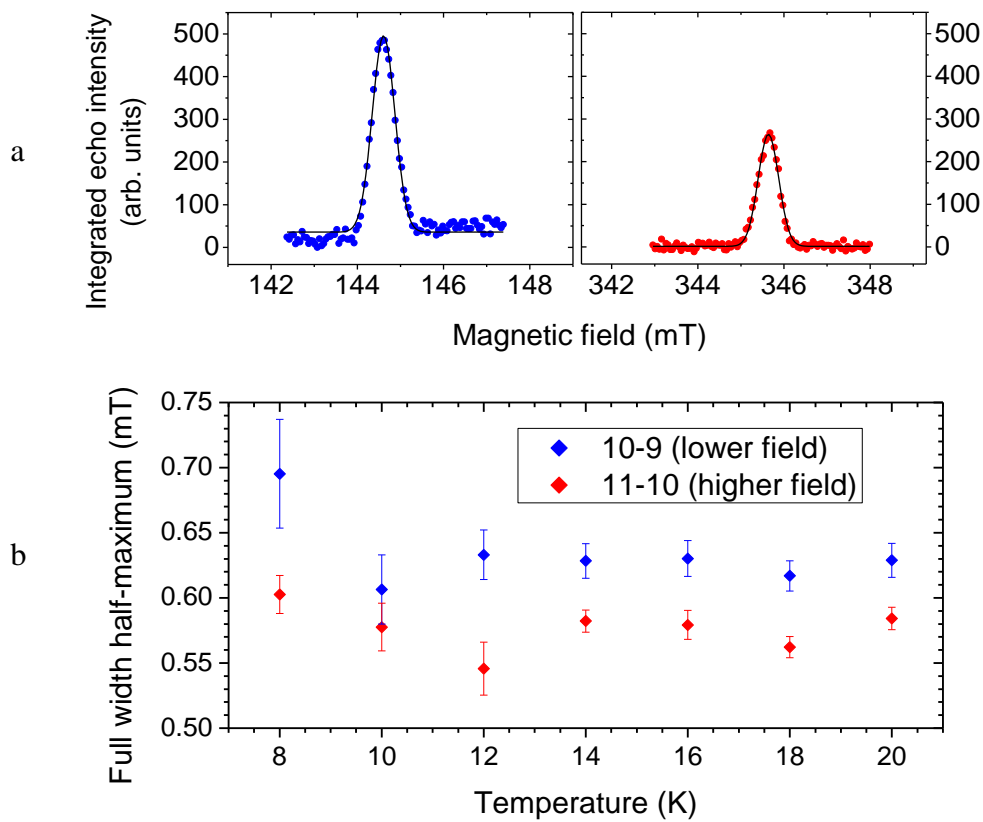

**Figure S2. Echo-detected field sweeps of Si:Bi at 4 GHz.** a) Field-swept spectra of bismuth-doped silicon at 18 K recorded as the size of the spin echo. The data are fitted with Gaussians 0.61 and 0.56 mT wide for the low and high field lines respectively. The intensity of these lines is not in the same ratio as the CW measurements, because other parameters are important in an echo-detected field sweep. In particular, the electron spin $T_2$ time of the 10-9 transition is longer than that of the 11-10 transition at 18 K. The resonant magnetic fields are slightly different to our other spectra because the microwave frequency was slightly different here: 4.060 GHz. b) Linewidths for both resonances as a function of temperature measured with echo-detected field sweeps. The error bars show the standard errors.

# The hybrid nuclear-electronic regime

Any system with coupled electronic and nuclear spins has some mixing so that the electronic eigenstates have some component of the nuclear spin and vice-versa. In an EPR spectrum this may be visible as small shifts of the resonances, but $2I + 1$ resonances are still present as in the unmixed case. In the unmixed limit, the spacing of the resonances is equal to the hyperfine coupling, $A$, and deviates from this as the mixing increases.

The qualitatively different feature of the hybrid nuclear-electronic regime is that, for excitation frequencies below $A(I + S)$, there are no longer $2I + 1$ resonances, but a richer pattern that depends on the chosen excitation frequency. For Si:Bi we have $A(I + S) = 5A = 7.4$ GHz and Fig. S3 shows that below this frequency there can be unusual effects including resonances that appear at two distinct magnetic fields as well as maxima and minima in the resonant frequency as a function of magnetic field.

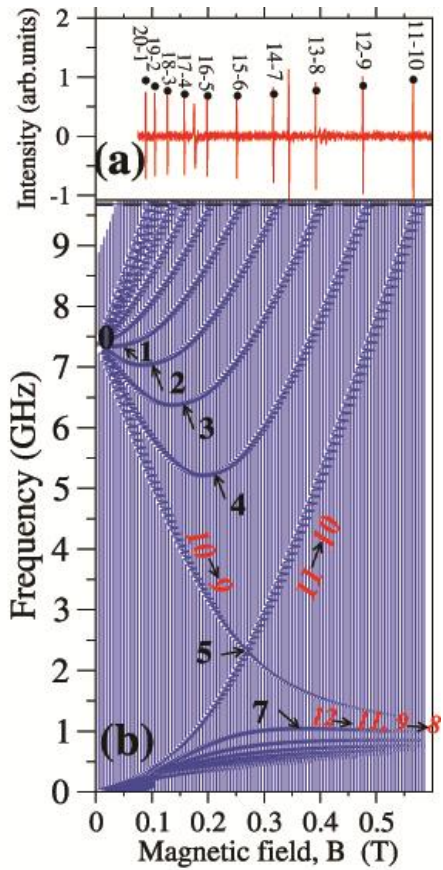

**Figure S3. The hybrid nuclear-electronic regime.** Si:Bi magnetic resonance as a function of applied magnetic field and excitation frequency (reproduced from our paper, reference 13, with permission from the APS). Previous experiments have used excitation above 7.4 GHz where ten resonances are observed, which smoothly shift to the ten high-field resonances for which electronic and nuclear spin are good quantum numbers[4]. a) EPR spectrum with 9.7 GHz excitation and a temperature of 42 K. b) Calculated magnetic resonance spectra.

## Decoherence measurements

The $T_2$ and $T_S$ times presented in Fig. 3b of the main text come from fits to the function $\exp(-t/T_2 - t^n/T_S^n)$. Fig. 3a and the 8 K data in Fig. S4a provide examples of these fits. In Fig. S4b, the exponent $n$ is plotted as a function of temperature. The values of $n$ are between 2 and 3 as with experiments[2-6] and theory[29,30] in the unhybridized regime.

The 10 GHz spin-lattice relaxation rates, $T_1^{-1}$ shown in Fig. 3b were fitted (in reference 4) with a power law (Raman) term added to an exponential (Orbach) term using the formula $T_1^{-1} = PT^7 + E\exp(-\Delta/k_B T)$ for temperature $T$ where $k_B$ is the Boltzmann constant. $\Delta$ was fixed at 500 K, while the other fitting parameters took the values (with standard errors) $P_{10\,\text{GHz}} = 1.26 \pm 0.03 \times 10^{-5}\,\text{s}^{-1}\text{K}^{-7}$, $E_{10\,\text{GHz}} = 3 \pm 0.1 \times 10^{12}\,\text{s}^{-1}$.

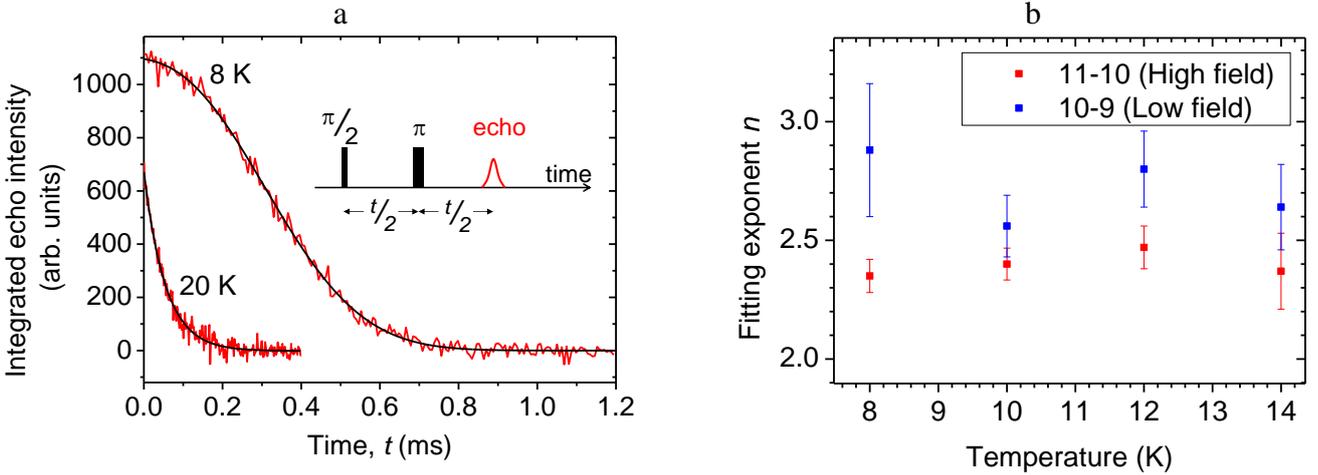

**Figure S4. Fitting Hahn echo decays.** The Hahn echo decays below 16 K are fitted with the function $\exp(-t/T_2 - t^n/T_S^n)$. a) The fit at 8 K for the 11-10 (higher field) resonance with the 20 K decay for comparison. b) Dependence of the fitting exponent $n$ on temperature.

## Decoherence simulations

In Fig. 3c of the main text the measured spin echo decays at 10 K are compared with our simulations using the cluster correlation expansion[30,S1,S2] using no free parameters. The good agreement shows that $^{29}$Si impurities dominate the spin echo decay.

A bismuth donor was surrounded by a bath of spin ½ $^{29}$Si nuclei. We calculated the spin echo decay of the bismuth due to dipolar interactions between pairs of $^{29}$Si nuclei without including decoherence due to phonons or interactions between donors. We assumed that the bath is not initially entangled with the bismuth donor. The bismuth donor is then put into an equal superposition of the two levels being excited: either the $|11\rangle$ and the $|10\rangle$ or the $|10\rangle$ and the $|9\rangle$. After a time $t/2$, a $\pi$ pulse is applied which is assumed to have negligible duration compared with $t$.

We construct the combined system density matrix and trace over the bath to obtain the bismuth density matrix. The modulus of the normalized off-diagonal element is proportional to the intensity of the spin echo. $^{29}$Si nuclei were placed at random sites in the diamond cubic lattice with an abundance of 0.0467. The donor is situated at the centre of the silicon cube, which has a side length that we varied from 5.2 nm to 31.6 nm. The reduced problem of the two-cluster bath was solved quantum mechanically for all pairs of $^{29}$Si spins with a maximum separation equal to either the 2$^{nd}$ or the 3$^{rd}$ nearest neighbour distance. For each spin pair, the reduced problem was solved separately for each of the four unentangled bath eigenstates, and the results averaged. In addition to the usual terms for the donor electron coupled to the bismuth nucleus[13] the two-cluster Hamiltonian included a pair of $^{29}$Si Zeeman terms, isotropic superhyperfine terms describing the interactions between the donor electron and each $^{29}$Si nucleus, and a dipolar term for the interaction between a pair of $^{29}$Si nuclei. The Fermi contact superhyperfine coupling was calculated from the Kohn-Luttinger electronic wavefunction for the bismuth dopant with 69 meV ionization energy in the silicon lattice. In obtaining the $^{29}$Si - $^{29}$Si dipolar terms it was assumed that the external magnetic field was large enough to conserve the total $^{29}$Si Zeeman energy. The two-cluster results were combined using the cluster correlation expansion[30]. Finally, we repeat this calculation for 100 random $^{29}$Si placements and report the average over these placements.

Figure S5a shows convergence of the two-cluster results as we increase the lattice size from 5.2 nm to 31.6 nm. These results also converge as the maximum separation between $^{29}$Si pairs is increased from the 2$^{nd}$ nearest neighbour to the 3$^{rd}$ nearest neighbour distance as shown in Fig. S5b. This gives us confidence in the results reported in Fig. 3c, where the agreement with experiment is good. The small discrepancies between theory and experiment are expected to be mainly due to ignoring alternative sources of decoherence in the simulation and limited knowledge of the donor electron wavefunction.

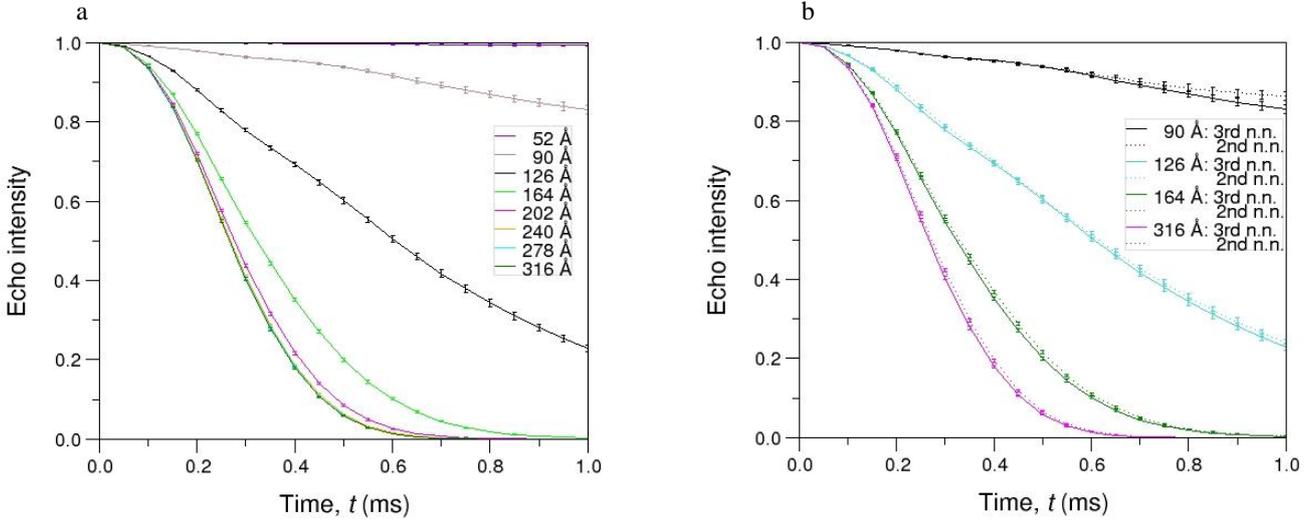

**Figure S5. Convergence of the two-cluster correlation expansion for spin echo decays in Si:Bi at 4 GHz.** a) Convergence as the lattice size is increased. Pairs of $^{29}$Si nuclei with separations up to the 3rd nearest neighbour distance in the silicon lattice were included in the calculation. b) Convergence as the maximum distance between paired $^{29}$Si nuclei is increased by pairing $2^{nd}$ and $3^{rd}$ nearest neighbours. The $2^{nd}$ and $3^{rd}$ nearest neighbour separations in the silicon lattice are $\frac{\sqrt{2}}{2}a_0$ and $\frac{\sqrt{11}}{4}a_0$ respectively, with $a_0$ = 0.543 nm. The results are compared for a range of lattice sizes. For both panels, the error bars are the standard deviation of the mean intensity after 100 random spatial configurations of $^{29}$Si nuclei, and the external magnetic field was chosen to be 0.3446 T so the 11-10 Si:Bi transition was excited.

We fitted the echo decay simulations from Fig. 3c with the function $\exp(-t/T_2 - t^n/T_S^n)$, as with the experimental measurements. For both the 11-10 and the 10-9 simulations, the $T_2$ is effectively infinite (over $10^{18}$ seconds), and the exponent $n$ is 2.27 with a standard error on the fit of ±0.01. The value of $T_S$ for the 11-10 transition (0.315 ms) is shorter than for the 10-9 transition (0.336 ms). This trend agrees with our experimental measurements, and we attribute this to the smaller gradient d$f$/d$B$ for the 10-9 transition.

## Entanglement

The entanglement between Si:Bi electronic and nuclear spins shown in Fig. 1a of the main paper is real: it is not pseudo-entanglement of the sort that has been studied in reference S3 or the early NMR quantum computing literature[12]. Indeed the amount of entanglement present in our experiment reaches a concurrence of $C$ = 0.92, which is over twice as large as the recent demonstration of $C$ = 0.43 using Si:P (ref. 16). However, the entanglement in our experiment is currently much less useful as a resource for quantum information technology. As with the NMR quantum computing literature, and references 16 and S3, the entangled partners cannot conveniently be separated so as to allow experiments such as teleportation[S4]. Additionally, for the entanglement present in our experiment to be useful for quantum computing, the entangled electronic and nuclear spins should be good quantum numbers. This would require moving the magnetic field from the entangled regime we study here to a higher field such as 1 T, without significant decoherence occurring. A magnetic field pulse of >$10^3$ T/s would be needed as the coherence times are around 1 ms. EPR has been performed in the past with magnetic field pulses of $10^4$ T/s (ref S5).

## Supplementary References